%% file: 0-main.tex
\documentclass[sn-mathphys-num]{sn-jnl}% Math and Physical Sciences Numbered Reference Style 
\usepackage{soul}%
\usepackage[inline]{enumitem}
\usepackage{graphicx}%
\usepackage{multirow}%
\usepackage{url}%
\usepackage{amsmath,amssymb,amsfonts}%
\usepackage{amsthm}%
\usepackage[title]{appendix}%
\usepackage{xcolor}%
\usepackage{textcomp}%
\usepackage{manyfoot}%
\usepackage{booktabs}%
\usepackage{algorithm}%
\usepackage{algorithmicx}%
\usepackage{algpseudocode}%
\usepackage{listings}%
\usepackage{mathptmx}%
\usepackage{newtxtext,newtxmath}%
\usepackage{subcaption}
\usepackage{comment}
\usepackage{float}

\definecolor{crema}{rgb}{0.953, 0.910, 0.820}

\usepackage{xcolor}
\newlength\MAX  

\begin{document}

\title[Telegram]{Mapping the Italian Telegram Ecosystem: Communities, Toxicity, and Hate Speech}

%Map of the Italian use of Telegram in 2023
%Map and characterization of the Italian use of Telegram in 2023
%Telegram Usage in Italy
%Mapping Telegram Usage in Italy
%Mapping and Analyzing Telegram Usage in Italy
%A Deep Dive into Telegram Usage in Italy
%Unpacking Telegram's Role in Italy
%The Italian Telegram Experience in 2023

% Esempi senza ITALY e senza 2023
%Insights into the Use and Structure of Telegram
%Mapping and Characterization of Telegram Usage
%Patterns and Dynamics of Telegram Use
%Insights into the Use and Structure of Telegram

\author[1,2]{\fnm{Lorenzo} \sur{Alvisi}}\email{lorenzo.alvisi@iit.cnr.it, lorenzo.alvisi@imtlucca.it}

\author*[2]{\fnm{Serena} \sur{Tardelli}}\email{serena.tardelli@iit.cnr.it}

\author[2]{\fnm{Maurizio} \sur{Tesconi}}\email{maurizio.tesconi@iit.cnr.it}

\affil[1]{\orgname{IMT School for Advanced Studies}, \orgaddress{\city{Lucca}, \country{Italy}}}

\affil[2]{\orgdiv{Institute of Informatics and Telematics}, \orgname{National Research Council of Italy}, \orgaddress{\city{Pisa}, \country{Italy}}}

%%==================================%%
%% Sample for unstructured abstract %%
%%==================================%%

\abstract{Telegram has become a major space for political discourse and alternative media. However, its lack of moderation allows misinformation, extremism, and toxicity to spread. While prior research focused on these particular phenomena or topics, these have mostly been examined separately, and a broader understanding of the Telegram ecosystem is still missing. In this work, we fill this gap by conducting a large-scale analysis of the Italian Telegram sphere, leveraging a dataset of 186 million messages from 13,151 chats collected in 2023. Using network analysis, Large Language Models, and toxicity detection tools, we examine how different thematic communities form, align ideologically, and engage in harmful discourse within the Italian cultural context. Results show strong thematic and ideological homophily. We also identify mixed ideological communities where far-left and far-right rhetoric coexist on particular geopolitical issues. Beyond political analysis, we find that toxicity, rather than being isolated in a few extreme chats, appears widely normalized within highly toxic communities. Moreover, we find that Italian discourse primarily targets Black people, Jews, and gay individuals independently of the topic.  Finally, we uncover common trend of intra-national hostility, where Italians often attack other Italians, reflecting regional and intra-regional cultural conflicts that can be traced back to old historical divisions. This study provides the first large-scale mapping of the Italian Telegram ecosystem, offering insights into ideological interactions, toxicity, and identity-targets of hate and contributing to research on online toxicity across different cultural and linguistic contexts on Telegram.} % \hl{linked to a particular country or language is still missing}
%Moreover, we find that Italian discourse primarily targets LGBTQ+ individuals and Black people, independently of the topic. 
%%================================%%
%% Sample for structured abstract %%
%%================================%%

% \abstract{\textbf{Purpose:} 
% \textbf{Methods:} 
% \textbf{Results:} 
% \textbf{Conclusion:} 

\keywords{Telegram, social network, communities, toxicity}

%%\pacs[JEL Classification]{D8, H51}

%%\pacs[MSC Classification]{35A01, 65L10, 65L12, 65L20, 65L70}
 
\maketitle

\input{1-introduction}

\input{2-related-work}

\input{3-dataset}

\input{4-method}
\input{5-results}

\input{5.1-RQ1-pol}
\input{5.2-RQ2-tox}

\input{5.3-RQ3-hate}

\input{6-discussion}

\input{7-conclusions}

\backmatter

%\bmhead{Supplementary information} If your article has accompanying supplementary file/s please state so here. 

\bmhead{Acknowledgements}
This work was partly supported by SoBigData.it which receives funding from European Union – NextGenerationEU – National Recovery and Resilience Plan (Piano Nazionale di Ripresa e Resilienza, PNRR) – Project: “SoBigData.it – Strengthening the Italian RI for Social Mining and Big Data Analytics” – Prot. IR0000013 – Avviso n. 3264 del 28/12/2021.; and by project SERICS (PE00000014) under the NRRP MUR program funded by the EU – NGEU.
During the preparation of this work the author(s) used OpenAI chatGPT in order to proof-check the grammar of some paragraphs and refine the language. After using this tool/service, the author(s) reviewed and edited the content as needed and take(s) full responsibility for the content of the publication.

\section*{Declarations}

\subsection*{Ethics approval and consent to participate}
Not applicable.

\subsection*{Consent for publication}
Not applicable.

\subsection*{Availability of data and materials}
To ensure reproducibility, we released an anonymized, privacy-preserving version of the network analysed during the current study, with the link to be provided upon acceptance.

\subsection*{Competing interests}
The authors declare that they have no competing interests.

\subsection*{Funding}
Not applicable.

\subsection*{Authors' contributions}
LA, ST, and MT conceived the research, interpreted the results, shaped the discussion, and reviewed the manuscript. LA, ST, and MT performed the research. LA collected and analyzed the data. LA, ST, and MT wrote the manuscript. ST and MT coordinated the research. All authors read and approved the final version of the manuscript.

%\bmhead{Acknowledgement}

\begin{appendices}

\section{Prompts Submitted to the LLMs}\label{secA1}

This appendix presents the prompts provided to the LLMs leveraged in this work. Tables report the prompt text as follows.

\input{tab-1-prompt-topic-mixtral}

\input{tab-2-prompt-political-chatGPT}

\input{tab-3-prompt-identity-attack-nemotron}

%%=============================================%%
%% For submissions to Nature Portfolio Journals %%
%% please use the heading ``Extended Data''.   %%
%%=============================================%%

%%=============================================================%%
%% Sample for another appendix section			       %%
%%=============================================================%%

%% \section{Example of another appendix section}\label{secA2}%
%% Appendices may be used for helpful, supporting or essential material that would otherwise 
%% clutter, break up or be distracting to the text. Appendices can consist of sections, figures, 
%% tables and equations etc.

\end{appendices}

%%===========================================================================================%%
%% If you are submitting to one of the Nature Portfolio journals, using the eJP submission   %%
%% system, please include the references within the manuscript file itself. You may do this  %%
%% by copying the reference list from your .bbl file, paste it into the main manuscript .tex %%
%% file, and delete the associated \verb+\bibliography+ commands.                            %%
%%===========================================================================================%%

\bibliography{bibliography}% common bib file

\end{document}

%% file: 1-introduction.tex
\section{Introduction}\label{section_introduction} 
Telegram has shown a substantial growth in recent years, becoming a key platform for digital communication. Its strong encryption, minimal content moderation and group-centric architecture have promoted a diverse ecosystem of discussions, from political debates to grassroots organizations. However, these same features have also allowed the unmoderated spread of disinformation, extremist content, and toxic discourse, raising concerns about ideological polarization and online harm.~\cite{scheffler2021telegram}.

Despite its growing relevance, Telegram remains largely unexplored with respect to other social media platforms like Twitter. Existing research on Telegram mostly focused on specific cases of harmful content, such as conspiracy~\cite{schulze2022far,hoseini2023globalization,alvisi2024unraveling}, far-right extremism~\cite{urman2022they,walther2021us}, and disinformation campaigns~\cite{ng2020analyzing,willaert2022disinformation,ng2024exploratory}. Moreover, scholarly attention has mostly focused on right-wing ideological spaces, with far-left clusters receiving comparatively little attention, as often considered less problematic or more difficult to identify\cite{walther2021us,zihiri2022qanon,urman2022they}.%harder to find 
In parallel, although toxic discourse is widely recognized as a central challenge in digital ecosystems, our understanding of its structural diffusion within Telegram communities remains limited. It is unclear whether toxicity is confined to a few extreme environments or broadly normalized across ideologically diverse discussions.

A key limitation in the study of Telegram is the scarcity of large-scale datasets. Most existing datasets are either incomplete due to the complexity of data collection or are focused on specific topics. In addition, a comprehensive mapping of a national or linguistic Telegram ecosystem is still missing.

In this work, we address these gaps by conducting a large-scale analysis of the Italian Telegram ecosystem. 
Building upon the methodology introduced in~\cite{alvisi2024unraveling}, and leveraging the forwarding mechanism as a proxy for homophily %interaction
we collected a dataset of over 186 million Italian-language messages posted throughout 2023 across $15,378$ public chats. This is, to the best of our knowledge, the most extensive dataset of Telegram messages with a clear spatial-linguistic focus that captures both large and small-scale conversations on a wide array of topics. %comprehensive
We build a directed weighted forwarding network and use several state-of-the-art Large Language Models (LLMs) to characterize individual chats in terms of most discussed topics and political orientation. Moreover, we assess message toxicity using the Perspective API and explore how hate speech is distributed in communities by identifying its main targets.

We formulate the following three questions:

\textbf{RQ1.} How do political orientations vary across Italian Telegram communities?

\textbf{RQ2.} Is toxicity concentrated in a few extreme chats, or is it widely distributed across communities?

\textbf{RQ3.} Who are the primary targets of hate speech, and how are they related to ideological alignments?

Our findings show strong thematic and ideological homophily. Politically oriented communities typically adhere to far-right or far-left narratives, but we also identified a hybrid ideological group in which left- and right-wing rhetorics coexist, especially with regard to geopolitical discourses. We find that toxicity is not limited to fringe groups but seems to be widely normalized within highly toxic communities. 
Finally, hate speech frequently targets LGBTQ+ and black people, but also reveals intra-regional antagonisms, in which Italians attack other Italian groups along old historical divisions.

Our findings suggest the need of context-specific moderation strategies and pave the way for future work on cross-linguistic comparisons and the social drivers of toxicity on Telegram.

%% file: 2-related-work.tex
\section{Related work}\label{section_related_work}

Although Telegram's user base and societal impact have expanded, studies on its wider ecosystem are still limited. Most studies focused on specific phenomena such as political extremism, misinformation, infodemic, and conspiracy, especially on other platforms~\cite{calamusa2020twitter,corti2022social,hashemi2023geographical,gambini2024anatomy}, leaving the overall thematic diversity and broader ecosystem of Telegram largely unexplored.

The majority of work on Telegram concentrated on specific phenomena such as extremism~\cite{bloom2019navigating,walther2021us} and political campaigns~\cite{vavryk2022mapping,castagna2023italian}. For instance, the authors in~\cite{urman2022they} analyzed the interconnections between extremist groups on Telegram, while authors in~\cite{bovet2022organization} analyzed conspiracy-related communities. In addition, comparative studies have examined how conspiracy narratives differ across language communities, revealing culturally specific patterns~\cite{alvisi2024unraveling}. Other research has targeted specific domains such as cryptocurrency-related fraud, uncovering cross-platform scam networks using Telegram as a central hub~\cite{nizzoli2020charting}. Indeed, the limited moderation of the platform and its privacy characteristics have been known to facilitate the proliferation of harmful content, including supremacist speech~\cite{guhl2020safe}, conspiracy theories~\cite{pasquetto2022disinformation}, and hate speech~\cite{vergani2022hate}.

While several large-scale Telegram datasets have been released~\cite{lamorgia2023tgdataset,baumgartner2020pushshift}, these typically center on specific issues or domain (e.g., extremism or misinformation), and do not offer focused analyses of national or linguistic Telegram spheres. To address these gaps, we present a large-scale analysis of the Italian Telegram ecosystem, investigating community structures, topic-driven clustering (homophily), network dynamics, ideological orientations, and toxicity patterns across both political and apolitical domains.

%Network analysis is considered a robust approach to understanding the structural and temporal evolution of online communities. 
Indeed, online toxicity has been extensively under scrutiny on platforms like Twitter, especially in political contexts~\cite{tardelli2024multifaceted, pierri2024drivers}, yet remains underexplored on Telegram and across various thematic domains. Recent work showed that toxic behavior tends to persist, and often intensify, within polarized online communities~\cite{avalle2024persistent, cavalini2024characterizing}, including Telegram. Moreover, research on deplatforming events showed that Telegram frequently serves as a refuge for banned extremist groups, subsequently increasing toxicity levels on the platform~\cite{wich2022introducing}. Despite these efforts, comprehensive knowledge regarding how toxicity spreads or evolves across different thematic settings is still limited. Existing research has largely focused on specific events or domains~\cite{blas2024unearthing, alvisi2024unraveling}, therefore leaving mostly unexplored larger patterns and relationships, a gap that we aim to address to understand how different thematic contexts influence the distribution of toxic discourse.

To the best of our knowledge, this is the first in-depth look at the Italian Telegram sphere, offering a broad overview of its landscape, main topics, political leanings, and levels of toxicity within its socio-cultural context.

%% file: 3-dataset.tex
\section{Data collection}\label{section_dataset}

\subsection{Terminology}
Telegram offers two main types of spaces for interaction: channels and groups. A channel is a one-to-many communication space where subscribers may watch the content but cannot participate personally, while only administrators may publish messages. They are used for broadcasting information to a large audience, such as news updates or public announcements, and also for sharing scams and other types of misleading content~\cite{nizzoli2020charting}. Conversely, groups facilitate many-to-many communication in which every member may contribute to the conversation. Recently, Telegram introduced linked chats, which allow a channel to be officially connected to an affiliated group, representing the same community. This connection is significant within the dynamics of Telegram, as content posted in a channel is automatically forwarded to the affiliated group, generating discussions in the community. 
Indeed, the \textit{forwarding} feature allows Telegram users to share messages between different chats, contributing to the dissemination of information. This feature is also fundamental for the snowball sampling technique used in this work, which is explained in the following sections.

In summary, we will use the term \textit{channel} to refer to a channel, \textit{linked chat} to indicate a the pair channel-affiliated group, and \textit{group} to denote a standalone group not associated with any channel. Additionally, we will employ the generic term \textit{chat} to broadly describe any of these categories—channels, linked chats, or groups—when distinctions among them are not relevant.
%il forwarding, che può essere fatto da utenti (o utenti amministratori di canali e quindi risulta un forward fatto da un canale ecccccc) non so se sarà utile inserirlo

\subsection{Initial Seed Collection of Public Telegram Chats}
We first retrieve a list of public Telegram chats to use as starting seed for our data collection. We retrieve them from two online sources: \textit{Telegram Italia}\footnote{www.telegramitalia.it}, a community-driven website cataloging Italian Telegram chats, and \textit{TGStat}\footnote{www.tgstat.com}, a website providing an extensive catalog of Telegram chats across various countries and languages. Both sources are widely used in literature for data collection~\cite{urman2022they,la2023sa,tikhomirova2021community}. Given our focus on the Italian-speaking messages, we retrieved chats classified under the Italian language from TGStat. Notably, unlike Telegram Italia, which includes both channels and groups, TGStat only provided channels classified in Italian.
As a result, we collected $20,229$ cataloged Italian chats from Telegram Italia and $1,241$ from TGStat, for a total of $21,470$ public Telegram chats, of which 14,767 consist of channels and 6,703 consist of groups. Neither platform provides information on whether there are linked chats among the channels and groups they list.

\subsection{Retrieving data from chat seeds}
We began by retrieving data posted in the $21,470$ chat seeds between January 1 and December 31, 2023. To ensure the reliability of the catalogs, we first verified whether the chats were indeed Italian, retaining only those that had at least $10$ comments in Italian and where the majority of messages were in Italian. Additionally, we discarded empty chats, chats whose content consisted solely of pictures, non-public chats, and chats that were inaccessible at the time of data collection. As a result, we were left with $4,360$ valid seeds out of the original $21,470$. Notably, TGStat yielded a higher percentage of valid seeds despite cataloging fewer Italian chats, with $63$\% ($776$ out of $1,241$) passing our checks. In contrast, only $18$\% ($3,584$ out of $20,229$) of the seeds from Telegram Italia were considered valid. Notably, this finding suggests the partial reliability and up-to-date accuracy of existing Telegram censuses.

\subsection{Categorization of Chat Seeds}\label{seed_categorizaion}
\begin{figure}[th!]
    \centering
    \includegraphics[width=0.8\textwidth]{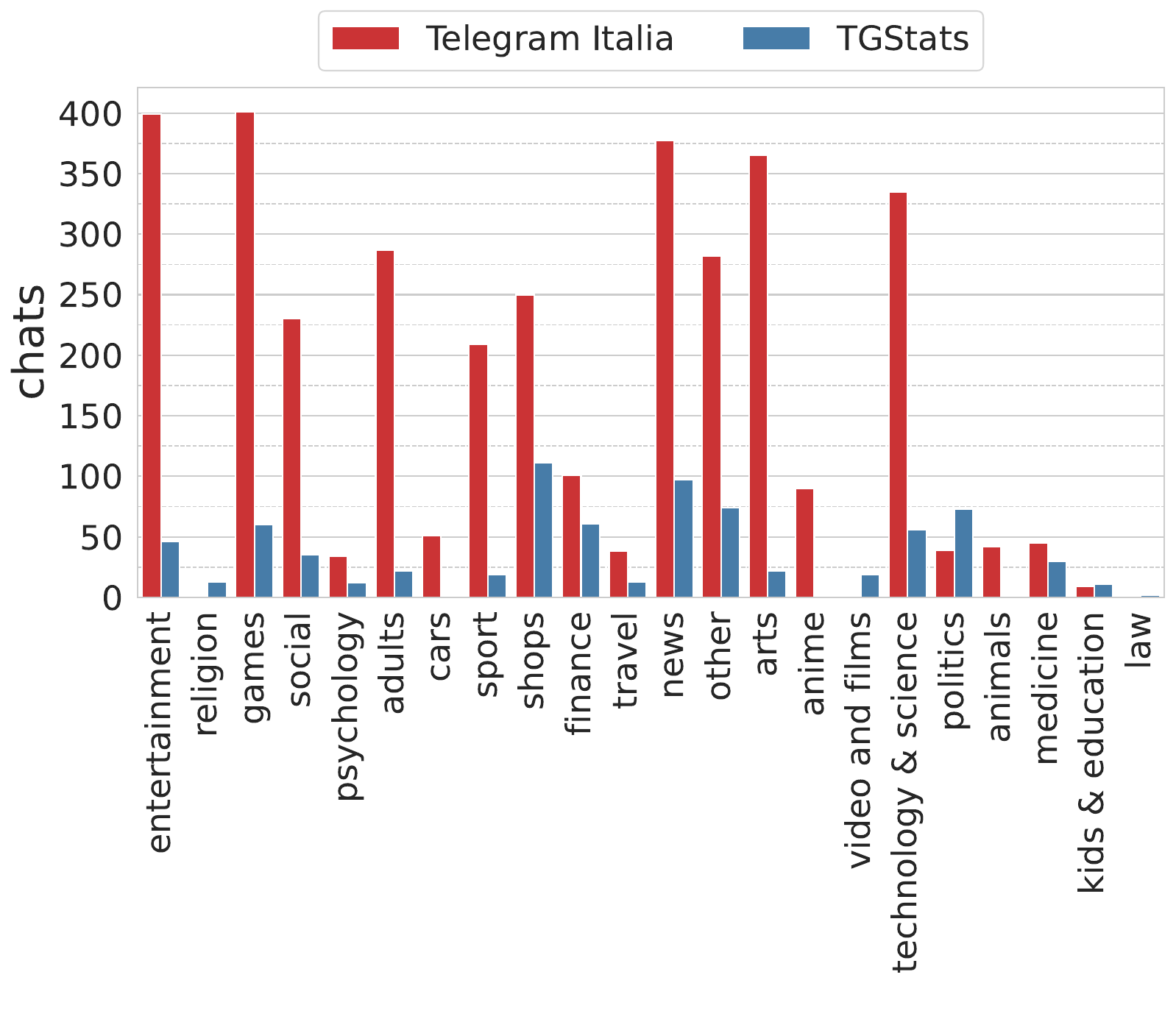}
    \caption{Count of chats per category for the $4,360$ valid seed chats as provided by their respective sources. The balance distribution across categories ensures a wide coverage of topics within the Italian Telegram ecosystem.}
    \label{fig:seed_categories}
\end{figure}

Both Telegram Italia and TGStat also provide a thematic classification for the chats in their catalogs. Telegram Italia uses category labels in Italian, while TGStat in English. Although the taxonomies differ in language, they are broadly comparable. For consistency, we aligned the categories across the two platforms using English labels. The categories are shown in Figure~\ref{fig:seed_categories}, which also illustrates the category distribution for the $4,360$ valid seed chats from which we successfully collected data. Interestingly, the distribution of chats across categories proved to be balanced, ensuring a wide coverage of topics within the Italian Telegram ecosystem.

\subsection{Data Collection via Snowball Sampling Strategy}
\begin{figure}[t!]
    \centering
    \includegraphics[width=0.8\textwidth]{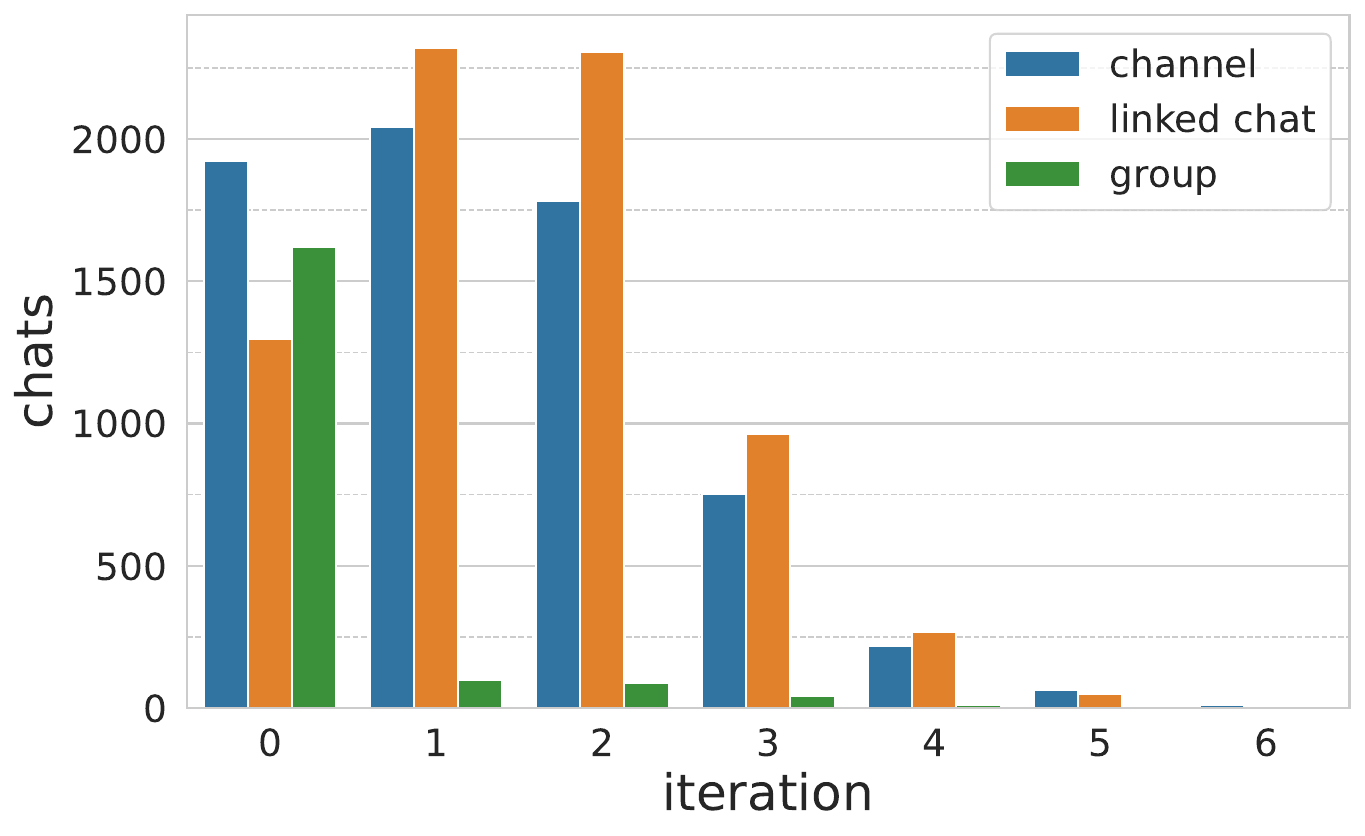}
    \caption{Counts and types of Telegram chats retrieved at each iteration of our snowball crawling strategy. Linked chats--consisting of a channel and its affiliated group--are the most prevalent, while standalone groups are the least represented in the dataset. This is likely because groups, by nature, tend to be more private, making them less accessible. }
    \label{fig:seed_downloaded}
\end{figure}

Our goal is to conduct the largest-ever census of all Italian Telegram chats. Starting from the seed chats, we initiated data collection using a snowball sampling strategy. This approach focuses on the diffusion of forwarded messages--a Telegram feature that allows users to share messages from one chat to another. It has been widely used in the literature for data collection, and its diffusion patterns have already provided valuable insights into detecting homophily and common interests in online communities~\cite{alvisi2024unraveling}. Coupled with the balanced distribution of thematic categories, this method ensures a broad and diverse dataset.
Starting with our $4,360$ seed chats, we retrieved their 2023 message history and followed the forwarded messages to discover new channels and crawl their messages. The process was repeated iteratively in a Dijkstra-style snowball fashion, where at each iteration, we retained only those chats containing at least $10$ messages and whose majority language was Italian. This process continued until no additional chats could be identified for download. Seven iterations were needed to exhaust the search. 
A the end of this process, we ended up with an unprecedented Italian Telegram dataset, consisting of $15,378$ chats and $186,757,857$ messages posted throughout 2023. With this comprehensive dataset, we are able to link channels with their affiliated groups, treating them as a single entity, which we will refer to as linked chats moving forward. In Figure~\ref{fig:seed_downloaded}, we present the count of chat downloads in each iteration, which follows a gamma distribution, reflecting the diminishing returns of each subsequent iteration.

%% file: 4-method.tex
\section{Methods}\label{section_method}
\subsection{Building the network}\label{sec: graph creation} 

The diffusion of forwarded messages plays a crucial role in shaping the structure of Telegram communities, as it highlights shared themes and common interests among different chats. This phenomenon reflects the principle of homophily — the tendency of individuals to connect with others who are similar to themselves — which becomes visible through the forwarding network~\cite{alvisi2024unraveling}.

To investigate the structure of the Italian Telegram sphere, we constructed a forwarding network modeled as a directed, weighted graph $\mathcal{G}:=(\mathcal{N},\mathcal{E})$. Each node represents a Telegram chat, which can be a group, a channel, or a linked chat. For every pair of nodes ${n,m} \in \mathcal{N}$, we draw a directed edge from $n$ to $m$ whenever a message is forwarded from $n$ to $m$, assigning to the edge a weight corresponding to the number of forwards. Linked chats — channels and their associated groups — are normalized by using the channel's identifier to represent the entire linked entity. Additionally, we remove self-loops to maintain consistency.

This process initially produced a network composed of $15,378$ nodes and $238,958$ edges. Although our aim was to comprehensively map the Italian public sphere on Telegram, the method was inherently limited by the coverage of the initial dataset (i.e., the seed list). Consequently, some segments of the Italian Telegram ecosystem may remain outside our scope. Nonetheless, chats disconnected from the giant component — those that neither received nor sent forwards — are likely to be less engaged or influential within the broader network.

To refine the network, we first excluded nodes without incoming or outgoing edges, reducing the graph to $13,144$ nodes. We then applied a fine-tuned version of the Louvain algorithm~\cite{dugue2015directed}, adapted for weighted and directed graphs, to detect communities. We retained only the communities representing at least $1$\% of the total nodes (i.e., communities with at least $132$ chats), resulting in a final network composed of $11,305$ nodes, $227,642$ edges, and organized into 15 distinct communities. %A visualization of the final network, including community labels discussed in the following sections, is presented in Figure~\ref{fig:network}.

\subsection{Topic labeling}
To identify the predominant topics discussed within each community, we employ Mixtral:8x7B~\cite{jiang2024mixtral}, an open-source sparse mixture-of-experts language model specifically designed to efficiently handle complex classification and generation tasks.

We assign a topic label to each chat using a predefined set of categories derived from the classifications provided by TGStat and Telegram Italia, as illustrated in Figure~\ref{fig:seed_categories}. The labeling process is guided by the prompt detailed in Table~\ref{tab:telegram_categorization}.
For each chat, we combine three sources of information to maximize the reliability of the labeling process:
\begin{enumerate*}[label=(\roman*)]
    \item the chat name,
    \item the chat description,
    \item and a random sample of messages posted in the chat.
\end{enumerate*}
Messages are randomly extracted until reaching a maximum of 4,000 characters - to balance the trade-off between execution time, performance, and hallucination probability — ensuring a representative snapshot of the chat's content without exceeding the model's constraints. Using multiple sources (i.e., name, description, messages) allows the model to better infer the dominant topic, especially for chats whose name or description alone might be ambiguous or sparse.
In this way, we assign a topic at the chat level based on the model's classification.

At the community level, we determine the dominant topic by aggregating the labels assigned to individual chats within each community and selecting the majority topic.

\subsection{Political labeling}
Shared political views are a significant factor driving homophily in online environments \cite{ackland2014political,halberstam2016homophily}. As such, we analyze the political leaning of chats using ChatGPT-4o~\cite{OpenAI2024ChatGPT4o}.Several studies have successfully employed this methodology \cite{burnham2024stance,zhang2023investigating}.
We present the prompt used for this method in Table~\ref{tab:telegram_labelling},  In particular, we use seven political labels based on the methodology adopted by MediaBiasFactCheck\footnote{https://mediabiasfactcheck.com/methodology/}, a website that fact-checks the credibility of news sources widely used in the literature to classify political orientations~\cite{tardelli2024multifaceted,tardelli2024temporal}. These labels include: \textit{far-left}, \textit{left}, \textit{left-center}, \textit{center}, \textit{right-center}, \textit{right}, and \textit{far-right}. We use categorical labels rather than numerical ones, as LLMs are more adapt at interpreting the contextual meaning of labels, as shown in the literature~\cite{schick2020exploiting}.
Political labels are initially assigned at the individual chat level by providing chatGPT-4o: \begin{enumerate*}[label=(\roman*)]
    \item the chat name,
    \item the chat description,
    \item and a random sample of messages posted in the chat.
\end{enumerate*}
Messages are randomly extracted until reaching a maximum of 5,000 characters. A chat may or may not exhibit a political leaning.
Additionally, ChatGPT-4o can generate error messages described with a reason such as self-harm, hate speech, sexual content, or violence that further provides valuable insights for chat evaluation.
Finally, at the community level, we classify a community as political (i.e., with a political leaning) if at least half of its chats exhibit a political orientation, with the community's political leaning determined by the most common leaning found in its labeled chats.

\subsection{Toxicity detection}
Toxicity has become an important area of research in recent years~\cite{avalle2024persistent}. To assess the presence of toxic language in our chat messages, we leverage the Perspective API,\footnote{https://www.perspectiveapi.com/.} a tool for detecting toxic language in online discussions developed by the Jigsaw unit within Google~\cite{lees2022new}. 

According to the documentation, the Perspective API ``predicts the perceived impact a comment may have on a conversation by evaluating that comment across a range of emotional concepts (attributes)''. Perspective can score the following attributes: \textit{Toxicity}, \textit{Severe toxicity}, \textit{Insult}, \textit{Profanity}, \textit{Identity attack}, \textit{Threat}, and \textit{Sexually explicit}. The returned score represents a probability, between 0 and 1, where higher scores indicate a greater likelihood that a reader would perceive the comment as containing the given attribute.
For this analysis, we leverage the \textit{Toxicity} attribute, which is also the most commonly studied in existing literature~\cite{hua2020towards, saveski2021structure, pierri2024drivers}, defined as ``a rude, disrespectful, or unreasonable comment that is likely to make you leave a discussion.''\footnote{https://developers.perspectiveapi.com/s/about-the-api-attributes-and-languages}. 

We analyzed all textual messages using the API in September 2024, successfully obtaining toxicity scores for 88\% of messages. Notably, the API supports multiple languages, including Italian. Finally, following the API guidelines, we label a message as toxic if its toxicity score is greater than $0.7$.\footnote{https://developers.perspectiveapi.com/s/about-the-api-score}

\subsection{Hate detection}
We leverage the \textit{Identity Attack} attribute provided by the Perspective API to detect whether a message contains hateful or discriminatory content. According to the Perspective API, the \textit{Identity Attack} attribute is defined as: ``Negative or hateful comments targeting someone because of their identity.'' The API returns a probability score between $0$ and $1$, where higher scores correspond to a greater likelihood that the message contains an identity attack. Following the recommended guidelines, we classify a message as containing an identity attack if its score exceeds a threshold of $0.7$.

Since the Perspective API only provides a probability score without specifying the targeted group, we adopt a semi-supervised approach to extract and identify the specific targets of hate within each message. First, we compile a set of identity categories based on previous literature~\cite{nobata2016abusive, ashforth1989social}, which includes: \textit{nationality}, \textit{ideology}, \textit{ethnicity}, \textit{religion}, \textit{sexual orientation}, \textit{gender}, \textit{age group}, \textit{disability}, \textit{profession}, and \textit{social class}.

Subsequently, we employ the \texttt{llama-3.1-nemotron-70b-instruct} model~\cite{wang2024helpsteer2preference,parmar2024nemotron}, a publicly available instruction-tuned model developed by NVIDIA,\footnote{\url{https://build.nvidia.com/nvidia/llama-3_1-nemotron-70b-instruct}} to classify each message by assigning it to one of the predefined identity categories. Table~\ref{tab:identity_attack} presents the prompt used for this classification task.

Finally, since the prompt is unsupervised, we need to standardize the resulting outputs. In fact, the model often generates various terms that refer to the same identity group, leading to inconsistencies. For instance, when identifying ethnicity-related targets, Nemotron may use variations such as ``African American'' and ``Afro-American'' to describe the same group. Although these expressions are conceptually equivalent, their lexical variability introduces noise into the categorization process.
To address this discrepancies, we implement a mapping strategy using ChatGPT-4o~\cite{OpenAI2024ChatGPT4o} and aggregate semantically similar terms into unified, standardized identity categories. This approach improves the consistency and reliability of the characterization of toxic communication within the interactions. To reduce the impact of potential biases introduced by the model, especially in sensitive cases involving labels related to ethnicity, sexual orientation, or gender, we adopted labels sourced directly from  literature in the respective domains.

%% file: 5-results.tex
\section{Results}\label{section_results}

% FIGURA DELLA RETE
\begin{figure}[t!]
\centering
\includegraphics[width=\textwidth]{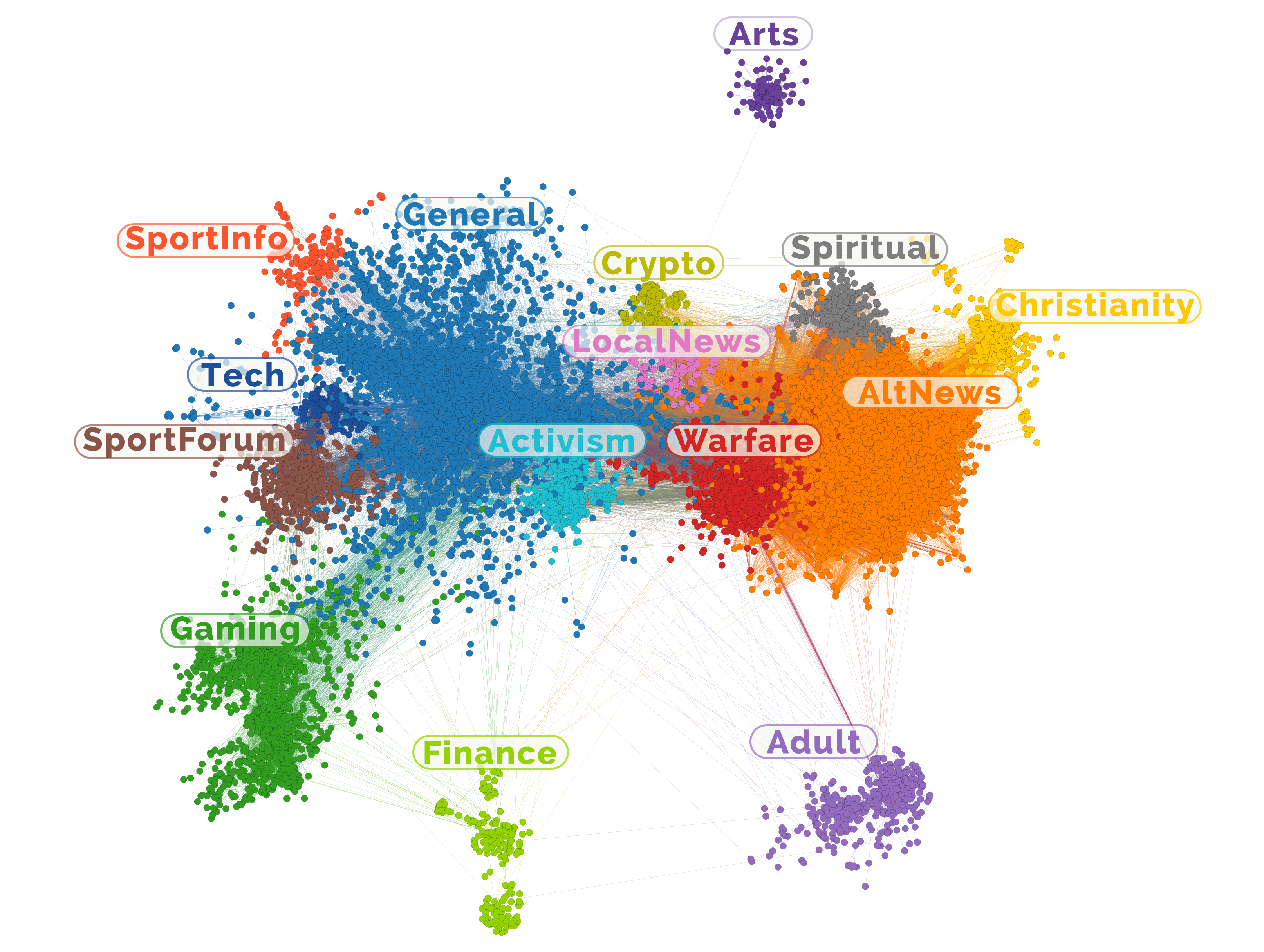}
\caption{Visualization of the chat network, colored by detected communities. Nodes represent chats; edges indicate forwarded messages. Clear community separation suggests distinct topical clusters and interaction patterns.}\label{fig:network}
\end{figure}

Applying our methodology, we obtained a forwarding network composed of $11,305$ nodes and $227,642$ edges, organized into 15 distinct communities. We identified these communities using a fine-tuned version of the Louvain algorithm~\cite{dugue2015directed}, specifically optimized for weighted and directed graphs, and focused our analysis on those communities comprising at least 1\% of the total nodes.

Figure~\ref{fig:network} illustrates the overall structure of the resulting network. To gain deeper insights into the nature of these communities, we first explore the main narratives discussed within their chats. We then further characterize each community along several dimensions, including political orientation, levels of toxicity, and primary targets of hate, addressing our research questions throughout the analysis.

\subsection{Narratives} 
Since our starting point is a set of seed chats, we rely on their associated thematic topics for categorization. To this end, we use Mixtral (as described in Section~\ref{section_method}) to assign topics to each chat, based on the predefined set of categories from TGStat and Telegram Italia (as described in Section~\ref{seed_categorizaion}). The topic of each community is then inferred based on the majority topic among its chats. 

The results are as follows: we identify two communities related to \textit{Entertainment},  three on \textit{News}, two on \textit{Sports}, two on \textit{Religion}, two on \textit{Finance}, one on \textit{Adult} content, composed mainly of adult-themed chat, one on \textit{Social} movements, referring to politically or culturally engaged communities, such as activist groups or grassroots organizations, rather than discussions about social media platforms, one on \textit{Tech} culture focused on tech-based shops, and one on \textit{Arts}, which appears to be associated with niche subcultures such as hikikomori, emo, introspectivity, love, and poetry, where artistic expression plays a key role in identity and self-representation. Interestingly, in the network representation, the \textit{Arts} community, while still connected, appears drifted away from the others, mirroring its introspective nature.  % This visual placement suggests that, much like its themes, the community exists on the edges, carving out its own space rather than integrating fully with the rest. % visually reflecting

Figure~\ref{fig:network} presents the communities renamed as follows. 
Given the limited scope of our predefined set of categories, some communities were assigned the same topic despite having subtle differences that our categories could not fully capture. Therefore, we compare the communities with the same category using a state-of-the-art method called Wordshift~\cite{gallagher2021generalized} and measure the differences in word usage across the chat descriptions of each community.

Results show that the two communities associated with the topic \textit{Entertainment}, while both falling under the same broad category, serve distinct purposes. One is centered around general mainstream discussions, making it the largest community in our dataset in terms of user base, chats, and comments. This community also appears to attract interactions from users across various social media platforms. Many popular Italian meme pages from Facebook, Instagram, and Twitter have a presence on Telegram within this community, as also observed during manual inspection. As such, we label this community as \texttt{General}. In contrast, the second community on \textit{Entertainment} focuses on gaming and online role-play conversations. For this reason, we name this community \texttt{Gaming}.
A similar distinction emerges in the three communities labeled with the topic \textit{News}, which each emphasize different aspects of information dissemination. One focuses on international news, with a strong emphasis on warfare, particularly the conflicts in Ukraine and Israel. We thus label this community as \texttt{Warfare}. The second community is about local news, either geographically based or centered around specific topics such as local job searches or climate forecasts, henceforth referred to as \texttt{LocalNews}. The third one is instead oriented toward personal freedom, counter-information, and alternative news, which we refer to as \texttt{AltNews}.

The two communities under the topic \textit{Sports} are primarily concerned with football, highlighting its central role in Italian culture and identity, with occasional mentions of other sports. However, they differ in their mode of interaction. One community mostly converge on discussion groups, while the other mostly comprises of channels where news, updates, and information exchange take front stage. As such, we label them as \texttt{SportForum} and \texttt{SportInfo}, respectively. 
Similarly, the two religious communities, though falling under the same broad category, diverge, as one reflects Italian cultural ideas through a focus on Christianity, prayers, saints, and the holy church, while the other emphasizes mystical activities and personal growth, spirituality and self-awakening. In light of these characteristics, we call these communities \texttt{Christianity} and \texttt{Spirituality}, respectively.
In addition, we observe a similar pattern for the communities annotated with the topic \textit{Finance}. In fact, one community is centered around cryptocurrencies, which we refer to as \texttt{Crypto}, while the other takes a broader approach, focusing on affiliate marketing and strategies for investments and generating passive income, which we refer to as \texttt{Finance}. 
Finally, we rename the community labeled with the topic \textit{Social} to \texttt{Activism}, to avoid confusion with social media-related content, as the term refers instead to activism, social centers, and grassroots engagement. %the broader and more ambiguous term.

Based on these distinctions, Table~\ref{tab:com} summarizes the communities and their topics.
The coherence of our categorization is further supported by the strong thematic alignment observed between each community's topic distribution and that of its seed chats.

\begin{table}[h]
\centering
\begin{tabular}{|l|l|r|r|}\hline
Name & Topic & Chats & Messages \\\hline
\texttt{General} & Entertainment &  3,763 &  71,719,216 \\
\texttt{AltNews} & News & 2,306 & 13,866,439 \\
\texttt{Gaming} & Entertainment & 1,235 & 7,266,662 \\
\texttt{Warfare} & News & 907 & 9,271,835 \\
\texttt{SportForum} & Sport & 485 &23,101,659  \\
\texttt{Adult} & Adult & 465 & 4,958,428 \\
\texttt{Spirituality} & Religion & 345 & 1,212,645 \\
\texttt{Activism} & Social & 334 & 856,596 \\
\texttt{Crypto} & Finance & 319 & 1,978,457 \\
\texttt{LocalNews} & News & 251 & 3,183,184 \\
\texttt{Christianity} & Religion & 240 & 497,578 \\
\texttt{Finance} & Finance & 184 & 1,620,185 \\
\texttt{SportInfo} & Sport & 151 & 6,371,009 \\
\texttt{Tech} & Tech & 151 & 4,820,753 \\
\texttt{Arts} & Arts & 139 & 378,093 \\\hline
\end{tabular}
\caption{Communities statistics}
\label{tab:com}
\end{table}

%% file: 5.1-RQ1-pol.tex
\subsection{Political Orientations (RQ1)} % of Italian Discussions

\begin{figure}[th] % mantenere [t] top, in modo che le figure siano sempre in cima alla pagine e non in mezzo al testo.
\centering
\includegraphics[width=\textwidth]{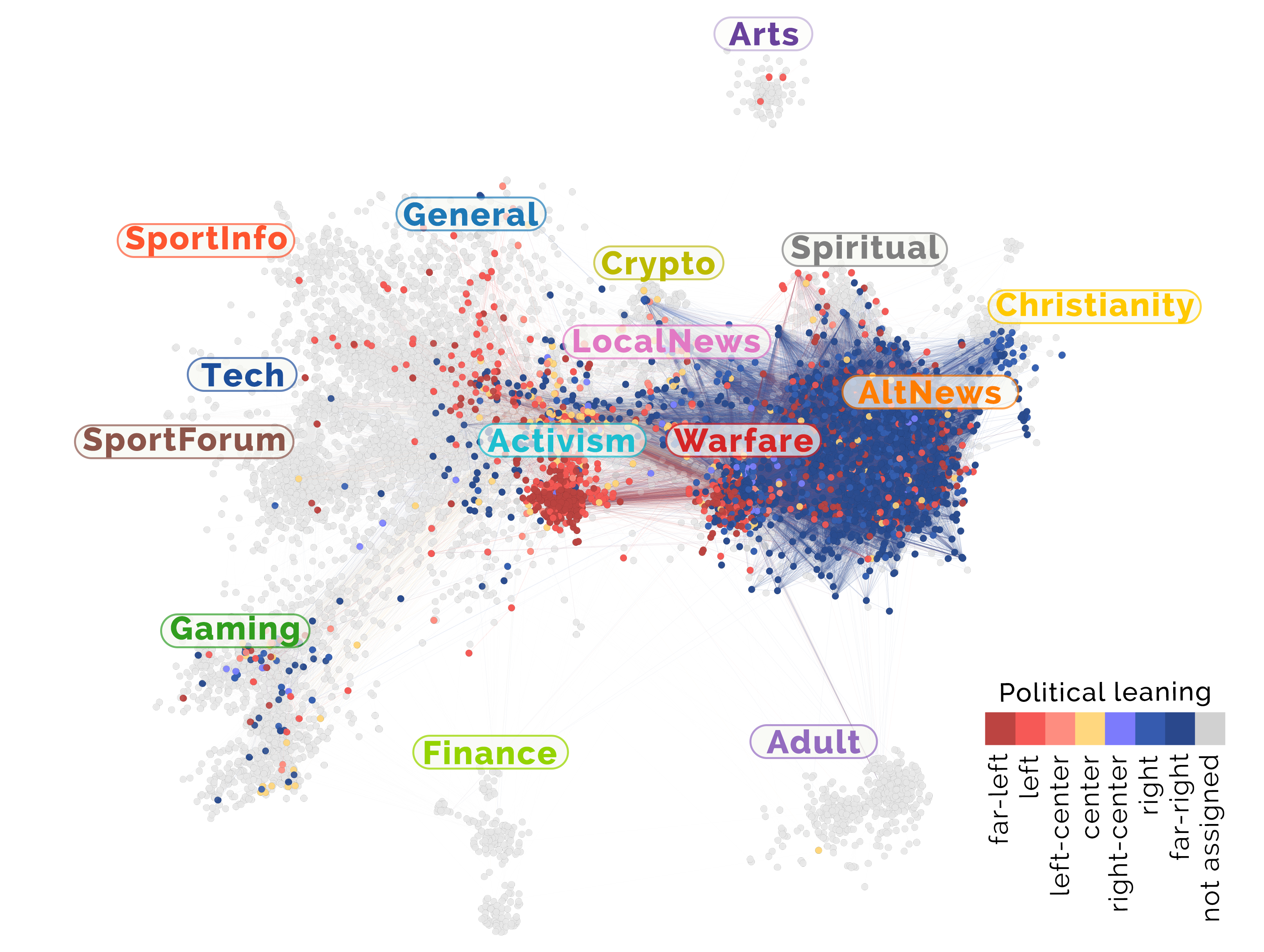}
\caption{Visualization of the chat network, colored by political leaning of each chat. Communities such as \texttt{Activism}, \texttt{AltNews}, and \texttt{Warfare} show clear ideological polarization, separating politically opposed groups.}\label{fig:network-pol}
\end{figure}

To answer (RQ1), i.e., how political leaning and the topics discussed within Telegram communities in the Italian public sphere are related, %whether and how Telegram communities within the Italian public sphere vary in their political leaning,
we first examine the distribution of political orientation across different communities. As described in Section~\ref{section_method}, a community is classified as political if at least 50\% of its chats exhibit a political orientation, with the dominant leaning determined by the most common label.

% ----------- Distribution of Political Leaning ----------- 
As shown in Figure~\ref{fig:network-pol}, most of the chats in our dataset do not show a clear political leaning. In fact, politically oriented chats are prevalent in five communities: \texttt{Activism}, \texttt{AltNews}, \texttt{Warfare}, \texttt{General}, and \texttt{Christianity}. Among these, only three (i.e., \texttt{AltNews}, \texttt{Warfare}, and \texttt{Activism}) meet the threshold for our definition political classification. In particular, the community \texttt{AltNews} is mainly composed of chats with a strong far-right political leaning, the community \texttt{Activism} is composed on mainly far-left chats, and \texttt{Warfare} contains both far-left and far-right political leaning chats. %%

% ----------- Political Polarization and Content Analysis ----------- 
The presence of both far-left and far-right narratives in \texttt{Warfare} reveals a complex ideological geopolitical landscape in which, for example, discussions on the Ukraine war involve both right-leaning chats typically expressing nationalist and anti-immigration ideas, and left-leaning chats matching with Russian rethoric and propaganda.
%, labeling Ukrainians as ``fascists.'' %% %This finding underscores the nuances of political discourse on Telegram, where ideological alignment does not always follow traditional categorizations. %This observation highlights how political categorizations in online discourse can be complex, %\hl{and sometimes counterintuitive}, warranting further discussion on the nuances of ideological alignment in digital spaces. %This bivalent nature is particularly notable in the context of discussions about
This observation highlights the complexity political views in online discourse, more than typically seen in studies of online discourse. As right-wing and extreme communities are typically connected with disinformation, radicalization, and social media moderation difficulties~\cite{schulze2022far}, research on online political debate really frequently focuses disproportionately on them. On the other hand, far-left internet networks get far less attention in the research. Our results, however, shows the existence of left-leaning groups on Telegram that occasionally participate in controversial debate, implying that a more all-encompassing strategy of researching political extremism in digital settings is required. %%

Although \texttt{General} and \texttt{Christianity} contain a substantial number of politically oriented chats ($442$ and $50$, respectively), these account for less than $50$\% of their total composition and thus do not qualify the communities as explicitly political. All remaining communities show low political content, with less than $20$ politically labeled chats and less than $10$\% of their composition associated with any ideological stance.

% ----------- Challenges in Political Classification ----------- 
%Finally, given that ChatGPT refused to label certain prompts containing content that violated its terms of use, we also examined the occurrence of classification errors. For instance, the adult-themed community, which shows no political leaning, had 300 prompts (64.5\% of its chats) labeled as sexually explicit. This suggests that while some topics, such as the differing stances on the Ukraine conflict, reflect political divides, others, like adult content discussions, are shaped by different forms of discourse regulation.

% ----------- Findings and Remarks ----------- 
\vspace{2ex}\noindent\textit{\textbf{Findings and Remarks}}\quad
Our analysis shows that political discourse is concentrated in a few communities, with clear ideological distinctions: \texttt{AltNews} is predominantly far-right, \texttt{Activism} is far-left, and \texttt{Warfare} is politically polarized, challenging the common focus on right-wing extremism in online political research, revealing that far-left communities also play a role in ideological discourse, sometimes engaging in contentious narratives.

These results suggest that while Telegram hosts a politically diverse ecosystem, the relationship between topic and political leaning is not straightforward, as users from various ideological backgrounds engage in a wide range of discussions across different topics.% Political Orientations and Topic Alignment meh

%% file: 5.2-RQ2-tox.tex
\subsection{Toxicity distribution across communities (RQ2)}
To answer RQ2, i.e., is community-level toxicity mainly driven by a few highly toxic chats, or does it result from toxicity being widely and evenly distributed across many chats within the community, %whether community toxicity is driven by a small number of highly toxic chats or by the contribution of many mildly toxic chats
we first analyze the distribution of toxicity across all chats. As mentioned, perspectiveAPI managed to successfully label $164$M messages, corresponding to the 88\% of our dataset. Following the official API guidelines, we classify a message as toxic if its toxicity score is greater or equal to $0.7$, resulting in a total of $2,3$M toxic messages in our dataset.
Next, we define toxicity at chat level by computing, for each chat, the percentage of toxic messages it contains. 
Finally, to assign a toxicity score to each community, we compute the average toxicity percentage of the chats within that community. 

\begin{figure}[th]
\centering
\begin{subfigure}{0.48\textwidth}
    \centering
    \includegraphics[width=\textwidth]{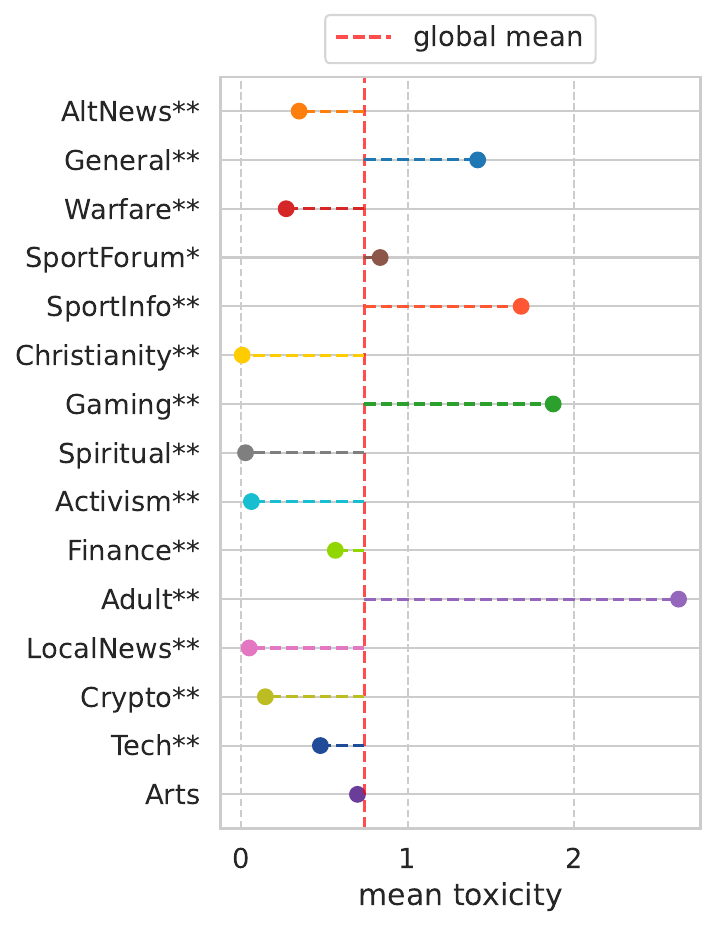}
    \caption{Mean toxicity percentages by community. Statistical significance levels indicate communities with notably higher or lower toxicity than average: $**: \rho < 0.01, *: \rho < 0.1.$}
    \label{fig:com-tox}
\end{subfigure}
\hfill
\begin{subfigure}{0.48\textwidth}
    \centering
    \includegraphics[width=\textwidth]{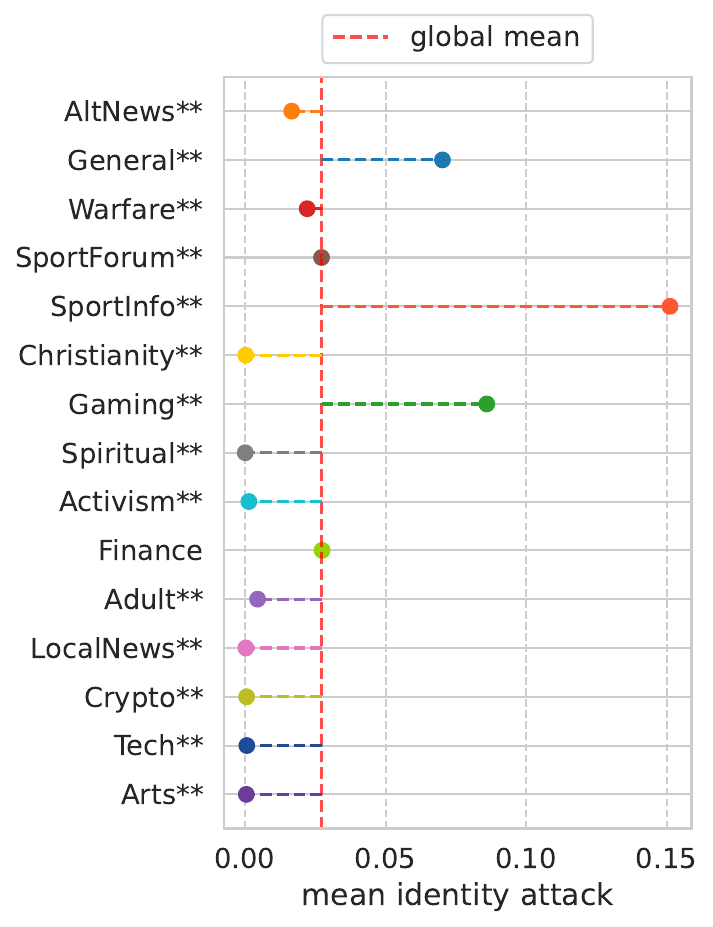}
    \caption{Mean identity attack percentages by community. Statistical significance levels indicate communities with notably higher or lower toxicity than average: $**: \rho < 0.01, *: \rho < 0.1.$}
    \label{fig:com-idattack}
\end{subfigure}
\caption{Toxicity and identity attack percentages by community.}
\label{fig:perspective-pallini}
\end{figure}

Figure~\ref{fig:com-tox} shows the mean percentages of toxicity across the communities. The toxicity of each community is represented by colored dots) 
Overall the toxicity of the communities spans from 0\% to  almost 3\%. In other words, on average, the chats within the communities contain between the 0\% and the 3\% of toxicity. While these percentages may initially appear low, it is important to consider that the dataset spans an entire year and includes conversations potentially containing millions of messages. Consequently, even small percentages translate into a significant number of toxic interactions.

We compared the average toxicity of each community against the overall average toxicity across all chats in the network, represented by the red line at $0.73$\%. This baseline indicates that the network, on average, exhibits relatively low toxicity. To determine whether each community significantly differs from this overall toxicity, we used the Mann-Whitney U test \cite{mann1947test}. This test was chosen as the distribution of the toxicity scores violates the assumptions of normality, and due to unequal sample sizes across communities. Specifically, we tested if the distribution of chat toxicity percentages within each community significantly differs from the toxicity percentages observed across all the chats in the network. This allowed us to identify communities with toxicity levels that are notably higher or lower than average. The results highlight statistical differences in toxicity levels across communities within the network.

As expected, the \texttt{Adult} community shows significantly higher toxicity levels compared to the overall average, confirming expectations about the nature of its content. Interestingly, the communities \texttt{General}, \texttt{Gaming}, and \texttt{SportInfo} also exhibit significantly higher toxicity levels, suggesting unexpected problematic interactions within these groups. Notably, topics of sports and entertainment have often been tied to antisocial behavior~\cite{kearns2023scoping} that, in extreme cases, can bring up to the normalization of toxicity~\cite{beres2021don}.
Conversely, communities such as \texttt{Tech}, \texttt{Crypto}, \texttt{AltNews}, \texttt{Warfare}, and \texttt{Activism} display significantly lower toxicity, indicating they might host more constructive or moderated discussions.
These findings highlight that toxicity is unevenly distributed across communities. %, with some areas that may require targeted moderation or deeper investigation into factors and topics driving their toxicity.
This suggests that some communities might require targeted moderation or further investigation to identify factors or topics contributing to their higher toxicity levels. % questa frase come link di intro a gini.

An important question arising from these observations is whether community toxicity mainly results from a few highly toxic chats or if it emerges from many chats, each contributing to the overall toxicity. Understanding this distinction can clarify how harmful behaviors spread and become normalized within communities. 
Additionally, this insight can support the design of more effective interventions aimed at promoting healthier interactions.
To answer this question, we used the Gini index, a measure commonly applied to study inequality or concentration within a distribution. In our analysis, the Gini index\cite{gini1912variabilita} measures how evenly toxicity is spread across the chats in each community. In particular, a high Gini index indicates that toxicity is concentrated in only a few chats, while a low Gini index means toxicity is more evenly shared among many chats.
We compared the mean toxicity percentage (indicating how toxic a community is on average) with the Gini index (indicating how toxicity is distributed within the community). Confronting these two measures allows us to understand if communities become toxic due to widespread normalization (i.e., low Gini) or because of a small number of highly problematic chats (i.e., high Gini). 

\begin{figure}[t!]
\centering
\centering
\includegraphics[width=0.8\textwidth]{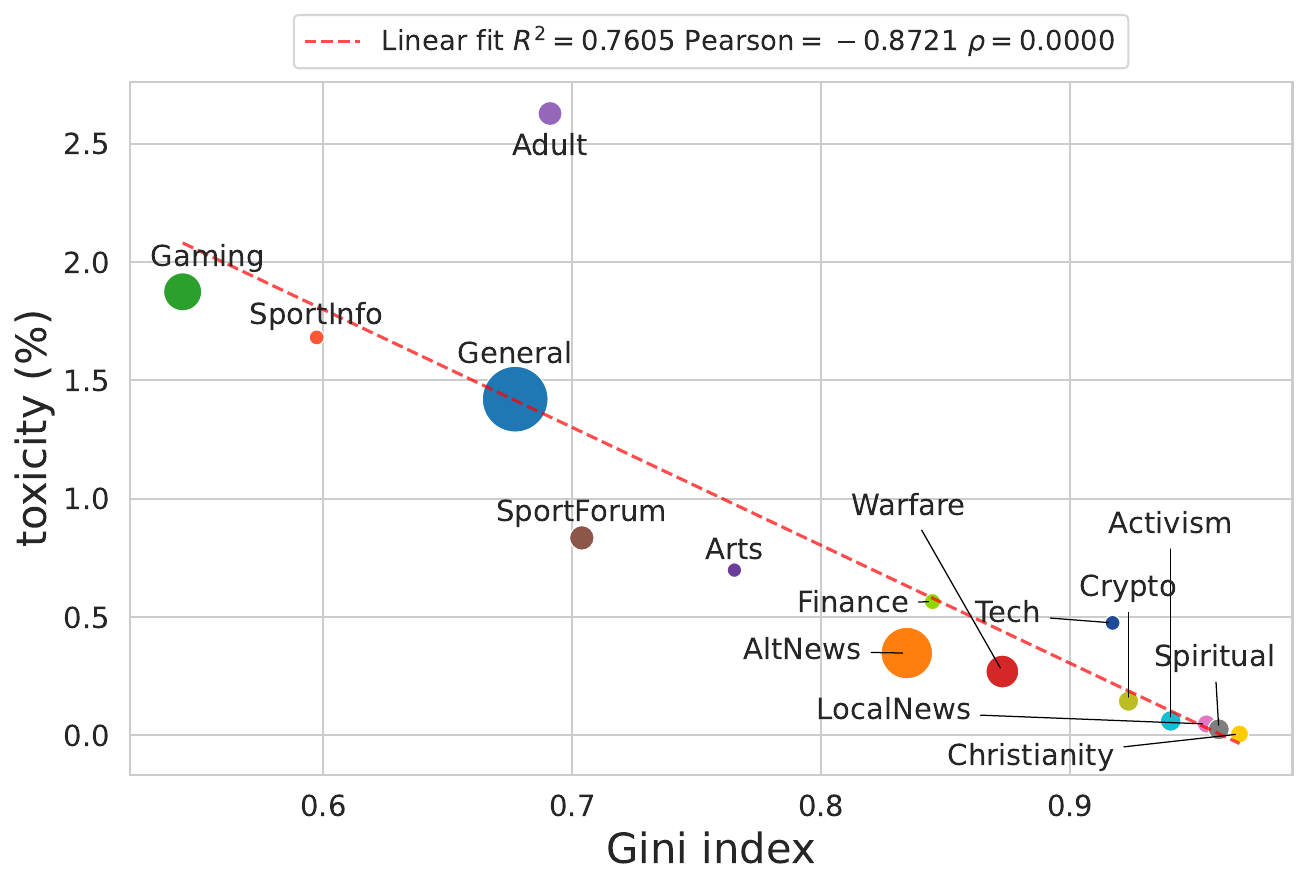}
\caption{Relationship between community mean toxicity and Gini index. Communities with higher toxicity generally have toxicity evenly distributed among chats, suggesting normalization of toxic behavior.}
\label{fig:gini-index}
\end{figure}

% Nella figura, il Gini index misura quanto è diseguale la distribuzione della tossicità all'interno di ciascuna comunità. Un Gini index basso (verso sinistra) indica che la tossicità è distribuita in modo più uniforme tra gli utenti o i messaggi della comunità. Un Gini index alto (verso destra) indica che la tossicità è concentrata in pochi utenti o messaggi. In questo grafico, si osserva che le comunità con tossicità media più alta tendono ad avere una distribuzione della tossicità più uniforme (Gini index più basso), suggerendo una normalizzazione del comportamento tossico.

Figure \ref{fig:gini-index} presents the relationship between the Gini index and mean toxicity percentage across communities. Each community is represented by a circle, with its size corresponfing to the number of chats it contains. The dashed red line shows a strong negative linear relationship between the Gini index and mean toxicity, supported by significant statistical evidence (Pearson's correlation $= -0.8721$, $R^2 = 0.7605$, $\rho < 0.0001$).
This result means that communities with higher average toxicity tend to have lower Gini indices, suggesting that toxicity is usually evenly spread across many chats. In other words, highly toxic communities typically become toxic because toxic behaviors become normalized across most of their chats, rather than being driven by only a few extremely toxic outliers. This linear dependency reinforces the idea that toxicity increases as it becomes normalized within a community~\cite{beres2021don}. 
The \texttt{Adult} community represents a special case: despite having an moderate Gini index, indicating that toxicity is moderately evenly spread across its chats, it still shows significantly higher average toxicity compared to other communities. This result likely reflects the explicit nature of adult discussions, where inherently sensitive or explicit topics naturally lead to higher overall levels of toxic interactions, independent of how concentrated toxicity is within specific chats.
Finally, community size does not appear related to either average toxicity or the Gini index. This means that community size alone does not significantly affect how toxicity develops or spreads. %influence

These insights can inform the design of effective moderation strategies. Communities with evenly spread toxicity (i.e., low Gini, high mean toxicity) likely require broader cultural or behavioral interventions, whereas communities with concentrated toxicity (i.e., high Gini, high mean toxicity) might benefit more from targeted interventions focused on specific problematic chats or users.

%% file: 5.3-RQ3-hate.tex
\subsection{Targets of Hate (RQ3)}
To answer RQ3, i.e., whether there are common targets of hate across Telegram communities and topics, we analyze the identity groups most frequently targeted by hateful messages. Using the Identity Attack attribute from the Perspective API and further classification with Llama-3.1-Nemotron-70B-Instruct, we examine attacks based on nationality, ethnicity, religion, sexual orientation, and gender.  As mentioned in Section~\ref{section_method}, we apply a mapping approach using ChatGPT-4o to standardize identity labels for consistency in all categories.
%% 

\begin{figure}[t!]
\centering
\includegraphics[width=0.8\textwidth]{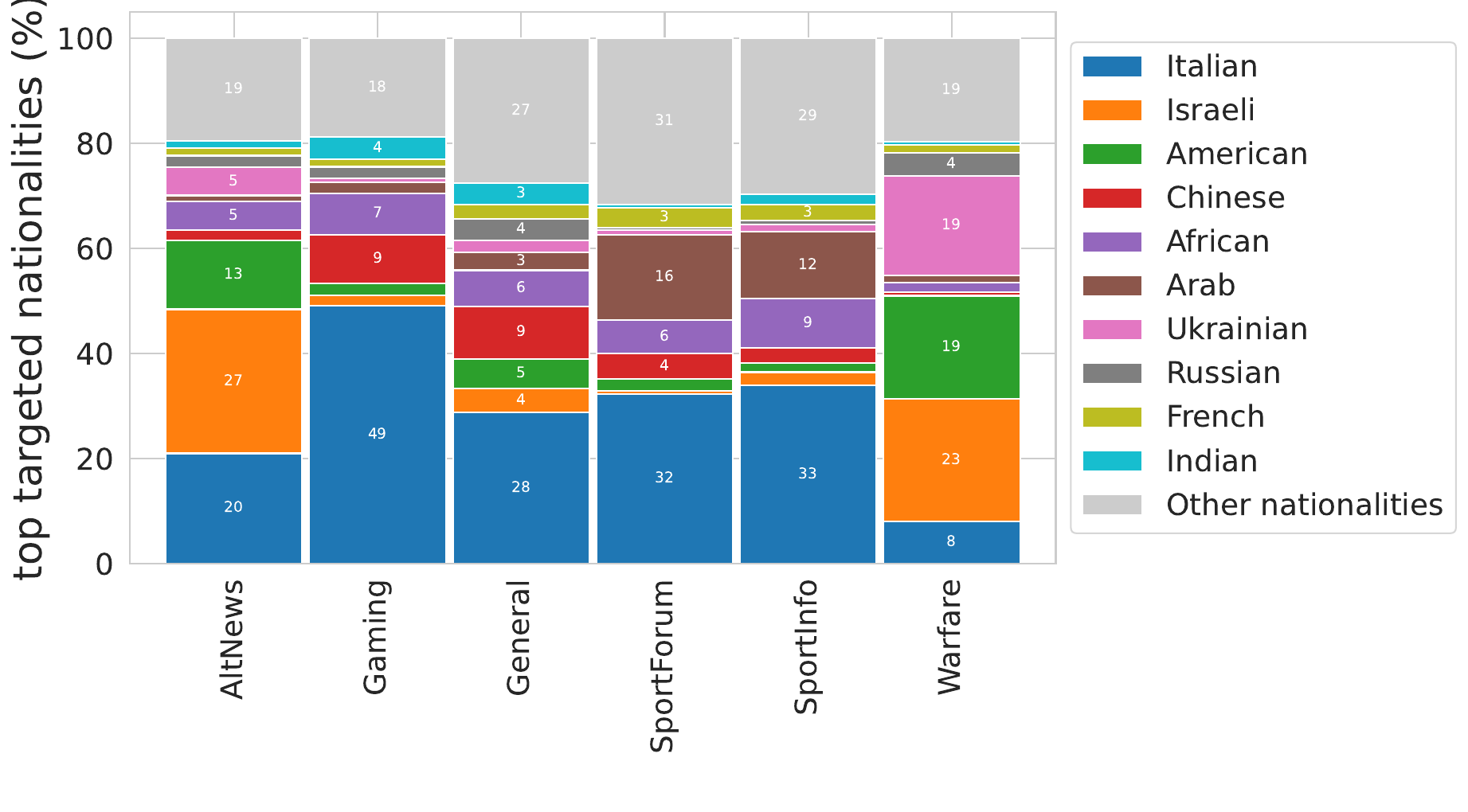}
\caption{Top targeted nationalities across communities. Italians are the most frequently targeted group, while Israeli, American, and Chinese users are also prominently targeted depending on the context. The order of nationalities in the legend follows their global frequency as targets.}
\label{fig:c1_nationality}
\end{figure}

%\paragraph{Hate Based on Nationality}
\vspace{2ex}\noindent\textit{\textbf{Hate Based on Nationality}}\quad
To analyze nationality-based attacks, we prompted the LLM to identify nationalities. In general, the model returned appropriate responses; however, in some instances, it produced broader regional or geographic labels (e.g., "Arab," "African") rather than specific nationalities. This reflects a known limitation of LLMs, which can be sensitive to linguistic context and occasionally inconsistent in how they represent complex social categories such as identity, culture, or origin. Despite these nuances, we retain the term nationalities for consistency. %  l’LLM ovviamente sbarella per i suoi motivi e quindi cattura aree geografiche
Figure~\ref{fig:c1_nationality} presents the targets on those communities that have at least 100 recognized identity attacks. Interestingly, Italians are frequently targeted across all communities. This trend highlights the cultural propensity in Italy for internal rivalries both regional and intra-national (e.g., parochialism, north vs south, etc.)~\cite{tak1988changing,villano2018competent}. Notably, these rivalries are present not only in sports-related communities, where they might be expected, but also in the other communities. 
Notably, hostility toward Israelis and Americans is most prominent in the \texttt{AltNews} and \texttt{Warfare} communities. Ukrainians also appear frequently in \texttt{Warfare}, reflecting broader political narratives in that context. Meanwhile, antagonism toward Arab nationalities is particularly pronounced within sports communities. 
Overall, while Italians appear as a cross-community target, most other nationalities tend to concentrate within specific thematic contexts, suggesting that nationality-based hostility is largely shaped by the dominant cultural, political, or ideological focus of each community. 

\vspace{2ex}\noindent\textit{\textbf{Hate Based on Ethnicity}}\quad
To analyze ethnicity-based attacks, we derive the categories for ethnicity from the classification proposed by the U.S. government agency~\cite{OMB1997race}. Notably, the Italian census adopts a different approach and classifies individuals based on their country of origin rather than ethnicity, reflecting a different cultural approach toward ethnic categorization~\cite{ISTAT2023ASI}. 
As such, to avoid repetition with the nationality-based analysis, we use the categories specified by the American government agency.

\begin{figure}[t]
\centering
\includegraphics[width=0.8\textwidth]{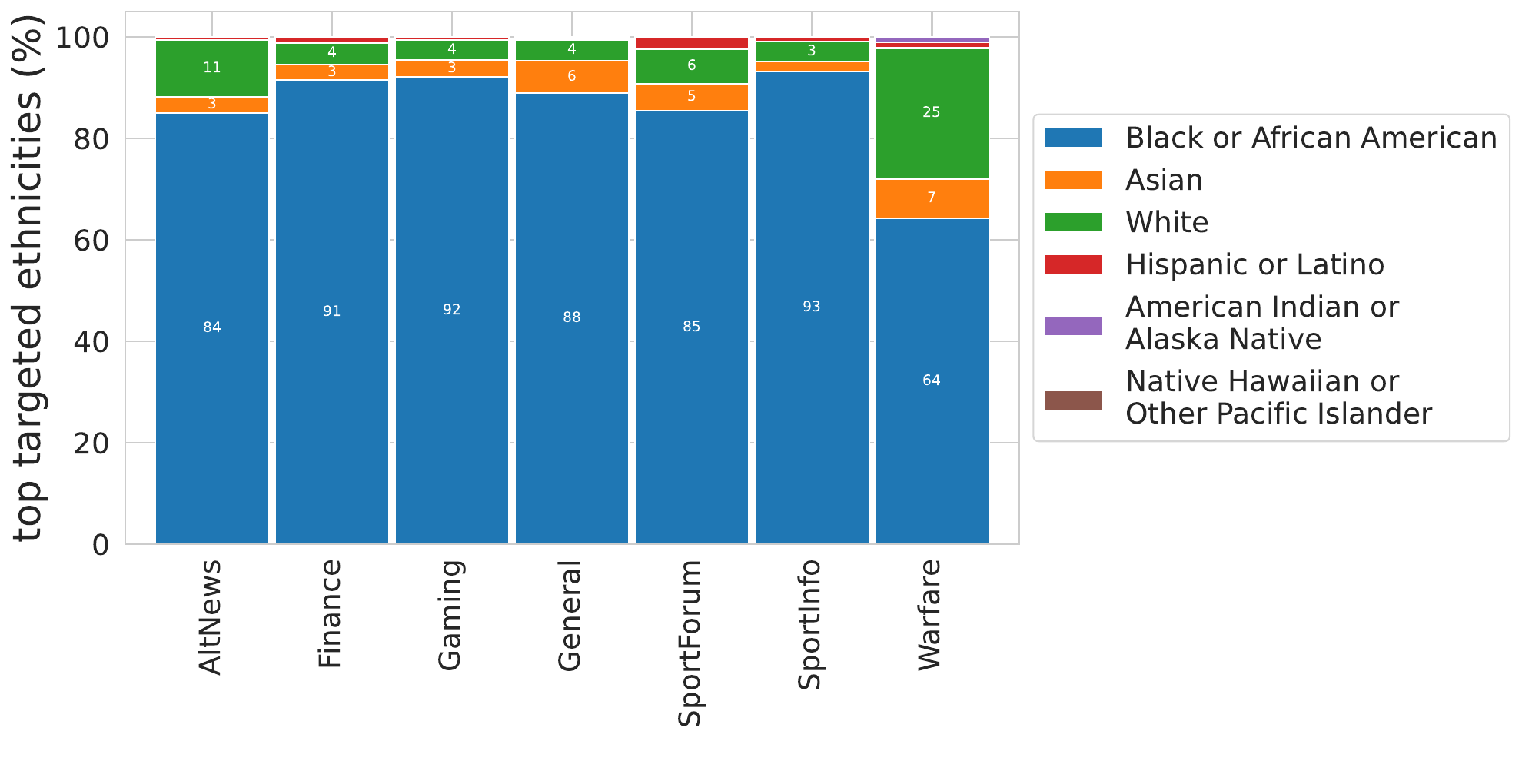}
\caption{Top targeted ethnic groups across communities.
Black or African Americans are the primary target in all communities, with notable increases in targeting of White and Asian users within the Warfare community. The order of ethnicities in the legend follows their global frequency as targets}
\label{fig:c2_ethnicity}
\end{figure}

Our results, presented in Figure~\ref{fig:c2_ethnicity}, reveal a clear trend: across all communities, Black and African American individuals consistently represent the primary targets of ethnicity-based hate. In most communities, between 85\% and 93\% of ethnicity-related attacks target Black individuals, indicating a widespread pattern of racial hostility that aligns with findings from existing literature~\cite{silva2016analyzing,chaudhry2020expressing}. A more heterogeneous distribution is observed in \texttt{Warfare}, where 64\% of attacks aimed at Black ethnicity is accompanied with 26\% of White targets. Notably, white group includes a wide spectrum of identities hl{(e.g., Slavic, Anglo, Israeli, Roma, etc.)}. %, therefore influencing interpretation and comparison across cultural contexts. %%%while also pointing to the challenges of interpreting broader ethnic categories across different cultural contexts.
Overall, the data highlights the centrality of anti-Black sentiment, while other targets appear more context-specific and harder to interpret without cultural nuance

\begin{figure}[t]
\centering
\includegraphics[width=0.8\textwidth]{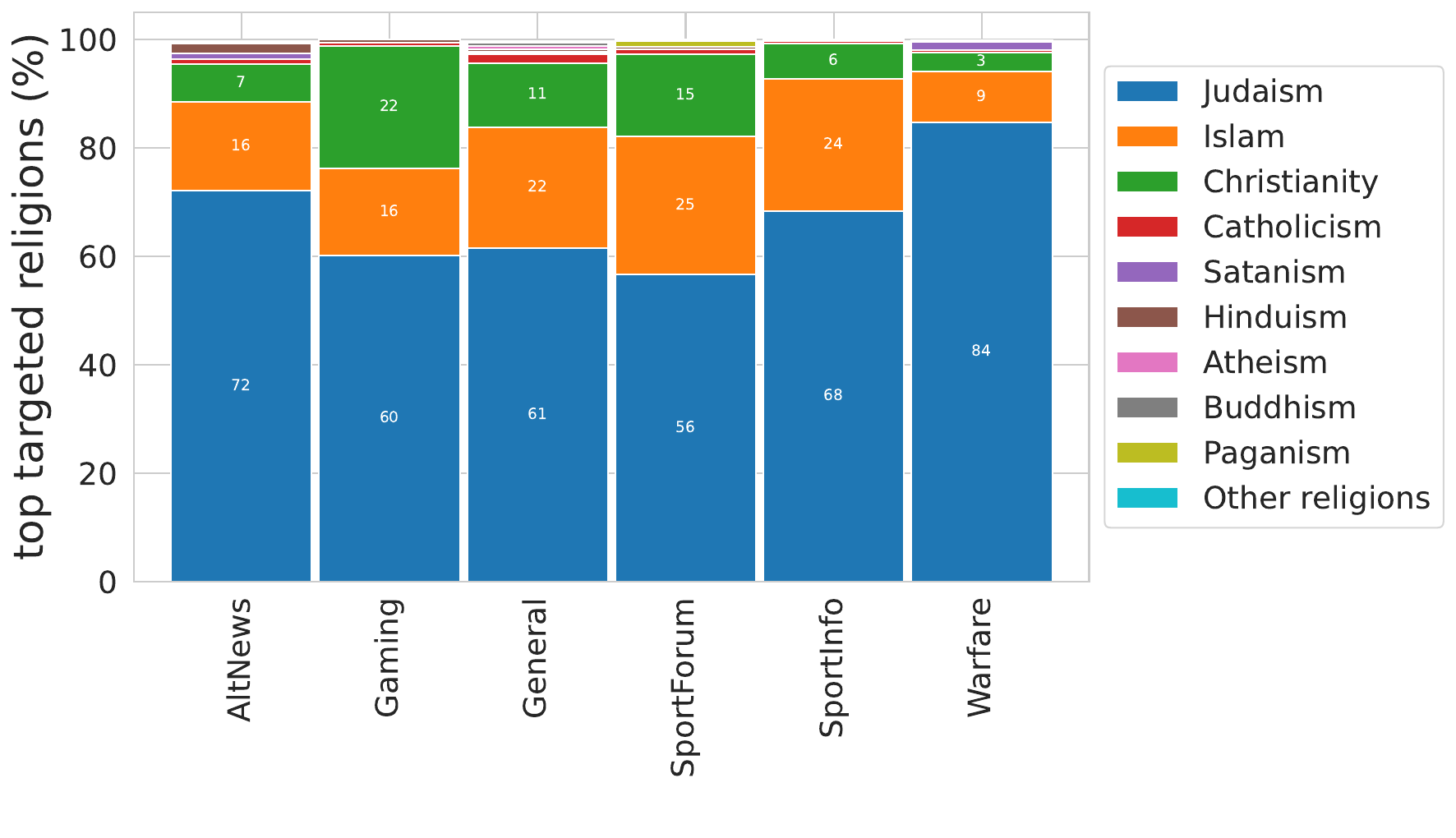}
\caption{Top targeted religious groups across communities.
Judaism is the most targeted religion in all communities, with Islam and Christianity also facing notable targeting—particularly in the sport and gaming communities. The order of religions in the legend follows their global frequency as targets}
\label{fig:c3_religion}
\end{figure}

\vspace{2ex}\noindent\textit{\textbf{Hate Based on Religion}}\quad
Results on religion-based hate (Figure~\ref{fig:c3_religion}) indicate that Judaism is the most frequently targeted religion across all communities, accounting for 57\% to 85\% of all religion-related hostility. Islam also receives 20\% of religion-related attacks in most communities. Other religious groups appear far less frequently, suggesting a concentration of antisemitic and islamophobic narratives. The results seams pretty stable across all communities, except in \texttt{Gaming} where Christianity is more attacked than Islam. However, it is important to note that our analysis is limited to data collected during 2023 and is thus influenced by the sociopolitical climate and major events of that period. For example, international conflicts, domestic political debates, or media coverage spikes may have shaped the salience of certain religious targets. As such, while the data reveals strong patterns, these should be interpreted within the broader temporal and discursive context in which they emerged.

\begin{figure}[t]
\centering
\includegraphics[width=0.8\textwidth]{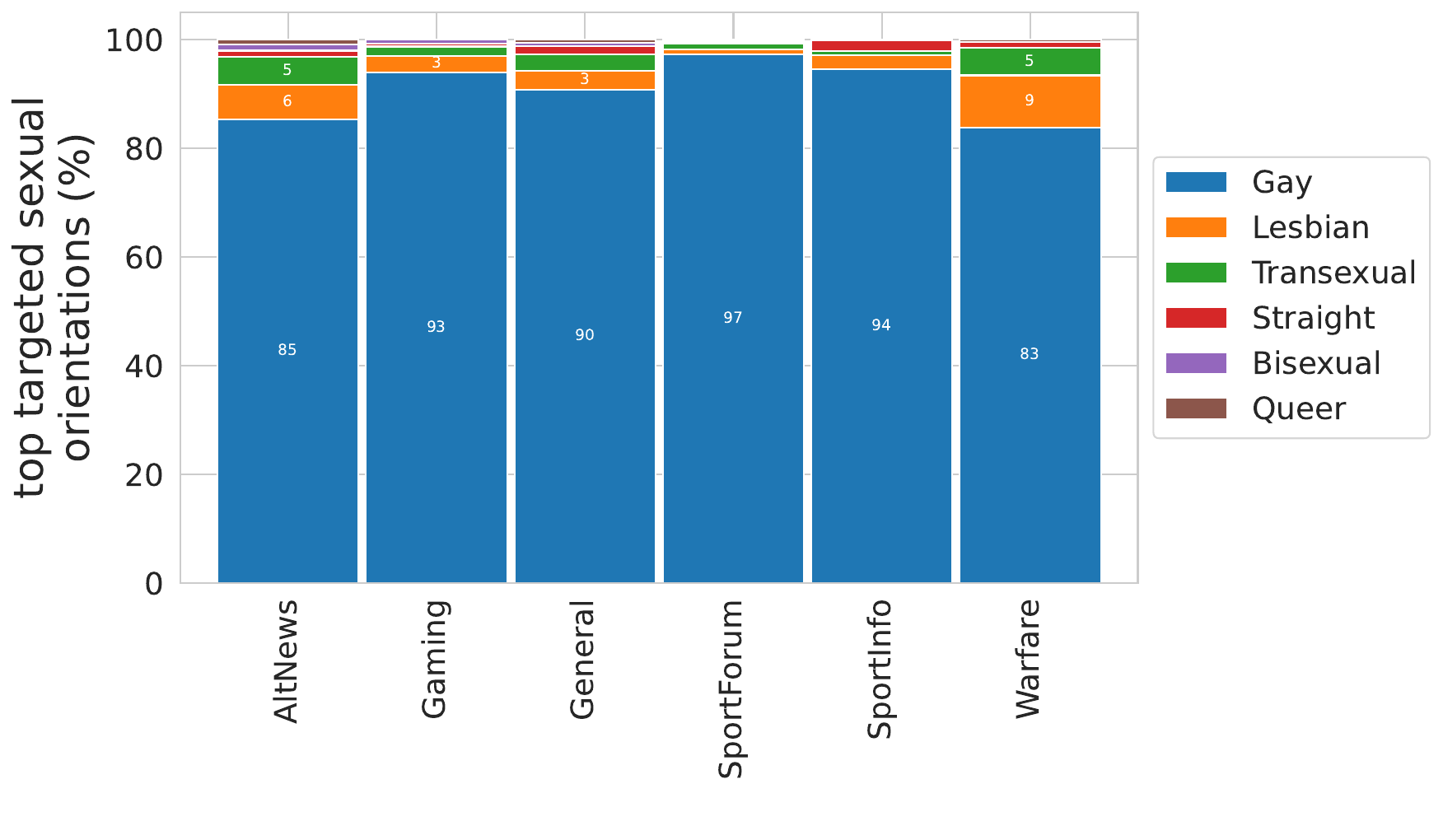}
\caption{Top targeted sexual orientations across communities.
Gay individuals appear as the primary target across all communities, although this may reflect a broader tendency to use the term generically to target the entire queer community. The order of sexual orientations in the legend follows their global frequency as targets}
\label{fig:c4_orientation}
\end{figure}

\vspace{2ex}\noindent\textit{\textbf{Hate Based on Sexual Orientation}}\quad
To standardize sexual orientation labels, we derive the labels from the LGBTQ+ acronym. As shown in Figure~\ref{fig:c4_orientation}, gay individuals are overwhelmingly the most targeted group, with over 85\% of sexual orientation-based attacks directed at them across all communities. Other identities within the LGBTQ+ spectrum, such as lesbian, bisexual, or transgender individuals,appear far less frequently in explicit attacks. This sharp imbalance may reflect both linguistic patterns in hate expression and broader social visibility dynamics: the term ``gay'' is often used in hate speech to target queer people in general, especially in online and meme-based discourse. Moreover, certain queer identities may be targeted through coded language, implicit hostility, or gendered slurs not captured directly by the sexual orientation categories used here. As such, these results should be interpreted as a conservative estimate of queerphobic discourse online.

\begin{figure}[t]
\centering
\includegraphics[width=0.8\textwidth]{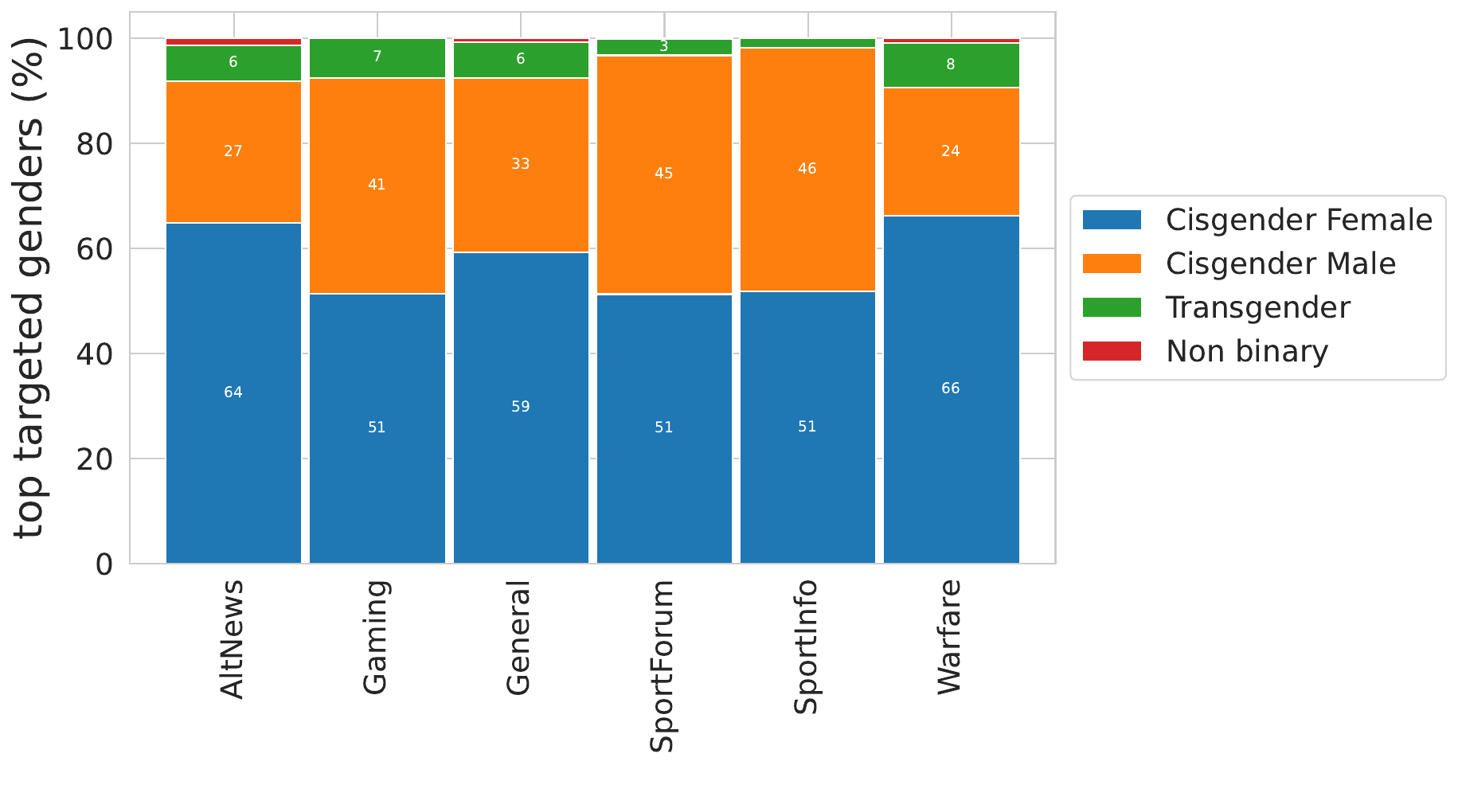}
\caption{Top targeted gender identities across communities. Cisgender females are the most targeted group overall, followed by cisgender males. The order of genders in the legend follows their global frequency as targets}
\label{fig:c5_gender}
\end{figure}

\vspace{2ex}\noindent\textit{\textbf{Hate Based on Gender}}\quad
For gender-based hate, we rely on terminology from existing gender discrimination studies~\cite{aparicio2018health, richards2016non}, which include both binary and non-binary gender identities. 
As shown in Figure~\ref{fig:c5_gender}, the majority of gender-related attacks are directed at cisgender women, followed by cisgender men. This trend is consistent with previous research on online misogyny, which highlights that women are more targeted in both political and non-political contexts~\cite{nadim2021silencing}. Moreover, while men also face a significant volume of online abuse, the nature of the harassment is often gendered, as women are more frequently subjected to sexualized and explicitly gendered attacks~\cite{erikson2023three}.

In contrast, transgender and non-binary identities appear far less frequently in our data. This relative scarcity may reflect their lower visibility within Telegram communities, or the use of indirect and coded language to target them, forms of hostility that are less likely to be captured through surface-level keyword analysis. Additionally, the strong presence of cultural and political narratives centered on traditional gender roles may shape which identities become focal points for attack.

Overall, while the distribution of gender-based hate is less fragmented than other identity categories, it reveals persistent patterns of gendered hostility, particularly toward women, that resonate with broader sociocultural dynamics across online platforms.

\vspace{2ex}\noindent\textit{\textbf{Findings and Remarks}}\quad
Our results show that hate speech in Italian Telegram communities follows clear patterns, with specific identity groups being targeted. Some forms of hate, like anti-Ukrainian sentiment in \texttt{Warfare} or hostility toward Arabs in sports discussions, depend on the context. However, hate against Italians, Black people, and gay people is widespread across all communities and independent from the topic. % certain prejudices are

%% file: 6-discussion.tex
\section{Discussions and Limitations}\label{section_discussion} %%

We analyzed the Italian Telegram landscape, showing that communities tend to form around ideological homophily. While most communities do not exhibit political orientation, we find strong political leanings in certain communities. In particular, some communities show both far-left and far-right rhetoric around geopolitical issues like the war in Ukraine, reflecting the broader trend in Italian politics where extremist views from opposite ends of the spectrum sometimes converge on specific topics. Moreover, by analyzing  toxicity levels across different communities, we show that online toxicity is not confined to ideological spaces, but is also present in entertainment, adult-themed, and sports communities. In particular, the highly emotional nature of sport fandoms and adult discussions promotes a competitive and more aggressive tone.
Our analysis of hate speech patterns further reinforces these findings. In fact, vulnerable minorities, such as LGBTQ+ individuals and Black people are the most frequently targeted identities across the communities, independently from the topic of the community. Instead, nationality-based hate varies by community ideology. In particular, both \texttt{AltNews} and \texttt{Warfare} communities present antisemitic discourse, with Israeli citizens being the most targeted nationality, aligning with the broader political narratives of these communities. Moreover, \texttt{AltNews} exhibits strong pro-Russian sentiment and strong hate directed at Ukrainians and Americans. Most importantly, results highlight a unique aspect of Italian online discourse, which is the widespread hostility directed at other Italians. This intra-national antagonism is rooted in historical regional divisions and cultural rivalries and emerges in all communities. While most evident in sports-related discussions, it is not confined to this domain. Political discussions also reflect this tendency, with users from different regions expressing resentment toward one another. This suggests that, beyond ideological or ethnic divides, internal cultural fragmentation plays an important role in shaping toxic interactions within the Italian Telegram ecosystem.

\subsection{Limitations}
While our study provides an unprecedented level of insight into the Italian Telegram ecosystem, it has certain limitations. In particular, the dataset contains public groups only and does not include private chats that might contain interesting insights into specific topics (e.g., invitation-only chats on cryptocurrency scam or pump-and-dump schemes~\cite{nizzoli2020charting}). As such, while our dataset is one of the most comprehensive Italian collections, it does not capture the entire Italian Telegram landscape. Nevertheless, we argue that the scale and scope of our dataset provide a highly representative sample of public discussions within the platform. Given the structure of Telegram, if a chat is not included in this dataset, it is likely because it operates in a closed or isolated manner, receiving little engagement from the broader network of public discussions.

%% file: 7-conclusions.tex
\section{Conclusions}\label{section_conclusions} %%

In this paper, we presented the most comprehensive analysis of the Italian Telegram ecosystem to date. We collected a large dataset of 186M messages from 13,151 chats. By employing a network-based approach, we mapped Italian Telegram communities. By using LLMs and the Perspective API, we characterized the thematic and political orientation of these communities and investigated the prevalence and distribution of toxic discourse and hate speech.
Our findings reveal that Telegram communities exhibit strong ideological homophily. In addition, most politically engaged communities aligning with either far-right or far-left narratives, although communities on geopolitical issues like the war in Ukraine show an ideological mix, blending far-left and far-right rhetoric. We also find that toxicity within highly toxic communities is widely spread among chats. Moreover Entertainment, adult and sports communities, which have no clear political stance, show very high levels of toxic content, suggesting that harmful discourse is not exclusive to ideological spaces. 
Additionally, our analysis of hate speech reveals that gay individuals, black people, and Jewish people are among the most frequently targeted groups. Moreover, we find that nationality-based hate varies depending on community ideology. However, broader Telegram discussions show high levels of intra-national (regional and local) hostility, with Italians frequently targeting other Italians. These findings show that hate speech on Telegram is driven by political views, topic, and cultural fragmentation, highlighting the need for context-specific moderation strategies. 
Finally, our dataset is limited to the Italian Telegram sphere, therefore findings may not be directly applicable to other linguistic or national contexts. Future works could explore how these patterns compare in English-language communities and investigate the factors that drive users toward toxic behavior.

%% file: tab-1-prompt-topic-mixtral.tex
\begin{table}[htbp]
    \centering
    \scriptsize
    \renewcommand{\arraystretch}{1.5}  % padding verticale
    \setlength{\tabcolsep}{6pt}  % padding orizzontale
    \begin{tabular}{|p{0.95\textwidth}|} 
        \hline
        \textbf{Prompt} \\ 
        \hline
        \textbf{Task:} \textit{Label Telegram channels} \\
        
        \textbf{Context:}  
        You will be provided with a sample of messages from a Telegram channel.  
        Your task is to assign the most appropriate label to the entire channel based on these messages, choosing from a fixed list of categories. \\

        \textbf{Input:}
        \begin{itemize}
            \scriptsize
            \item Fixed list of categories: \texttt{$<$CATEGORIES$>$}
            \item Chat name: \textit{$<$chat\_name$>$}
            \item Chat description: \textit{$<$chat\_description$>$}
            \item Sample messages: \textit{$<$messages$>$}
        \end{itemize}

        \textbf{Instructions:}
        \begin{enumerate}
            \scriptsize
            \item Familiarize yourself with the provided categories.
            \item Read all the messages from the channel.
            \item Analyze the overall content, tone, and purpose.
            \item Select up to 3 relevant categories, in order of importance.
            \item Provide the result in the specified JSON format.
        \end{enumerate}

        \textbf{Output Format:}
        \begin{itemize}
            \scriptsize
            \item \texttt{\{"categories": ["category1", "category2", "category3"]\}}
        \end{itemize}

        \textbf{Notes:}
        \begin{itemize}
            \scriptsize
            \item Select 1 to 3 categories in order of relevance.
            \item Do not create or merge categories.
            \item No explanations or extra information.
            \item Categories must exactly match those from \texttt{input1}.
        \end{itemize}

        \textbf{Example Output:}
        \begin{itemize}
            \scriptsize
            \item \texttt{\{"categories": ["anime", "arts", "games"]\}}
            \item If only two categories apply: \texttt{\{"categories": ["news", "politics"]\}}
        \end{itemize} \\
        \hline
    \end{tabular}
    \caption{Prompt for topic labeling of Telegram chats using Mixtral:8x7B}
    \label{tab:telegram_categorization}
\end{table}

%% file: tab-2-prompt-political-chatGPT.tex
\begin{table}[htbp]
    \scriptsize
    \centering
    \renewcommand{\arraystretch}{1.5}
    \setlength{\tabcolsep}{6pt}
    \begin{tabular}{|p{0.95\textwidth}|} 
        \hline
        \textbf{Prompt} \\ 
        \hline
        \textbf{Task:} Label Telegram channels \\

        \textbf{Context:} You will be provided with a sample of messages from a Telegram channel. Your task is to evaluate its political leaning. \\

        \textbf{Input:} \\
        
        The chat name: \textit{$<$chat\_name$>$}
        
        The chat description: \textit{$<$chat\_description$>$}
        
        A sample of messages from the Telegram channel: \textit{$<$messages$>$}
        
        Fixed list of labels: Far-right, right, right-center, center, left-center, left, far-left\\

        \textbf{Instructions:}
        \begin{enumerate}
            \scriptsize
            \item Read through all the provided messages from the channel.
            \item Analyze the overall content, tone, and purpose of the channel.
            \item Choose the political leaning of the channel and label it with one of the given labels.
            \item Output the result in the specified Python-readable JSON format.
        \end{enumerate}

        \textbf{Output Format:}
        Provide your response in the following JSON structure: \\

        \texttt{\{"political\_leaning": str | null\}} \\

        \textbf{Notes:} \\
        
        The key must always be "political\_leaning."
        
        The value must be one of the given labels.
        
        If you cannot label the political leaning from the given context, assign \texttt{null}.
        
        Do not create new categories or combine existing ones.
        
        Do not include any explanation or additional information.\\
        \hline
    \end{tabular}
    \caption{Prompt for political labeling of Telegram chats using ChatGPT-4o}
    \label{tab:telegram_labelling}
\end{table}

%% file: tab-3-prompt-identity-attack-nemotron.tex
\begin{table}[htbp]
    \scriptsize
    \centering
    \renewcommand{\arraystretch}{1.5}
    \setlength{\tabcolsep}{6pt}
    \begin{tabular}{|p{0.95\textwidth}|} 
        \hline
        \textbf{Prompt} \\ 
        \hline
        \textbf{Task:} Label Telegram message \\

        \textbf{Context:} You will be provided with a message from Telegram which contains an attack over one or more identities. Your task is to assign the most appropriate label to targeted individuals.\\

        \textbf{Input:}\\
        The message: \textit{$<$message$>$}\\

        \textbf{Instructions:}
        \begin{enumerate}
            \scriptsize
            \item Read the message.
            \item Identify the most appropriate attacked identity or identities. You need to specify the specific ethnicity, nationality, religion, gender, sexual orientation, disability, or other relevant identity characteristics of the victim(s) of the attack.
            \item Output the result in the specified JSON format.
        \end{enumerate}

        \textbf{Output Format:}
        Provide your response in the following JSON structure:\\

            \{\texttt{"targeted\_nationalities": list[str] | null},
            
            \texttt{"targeted\_ideologies": list[str] | null},
            
            \texttt{"targeted\_ethnicities": list[str] | null},
            
            \texttt{"targeted\_religions": list[str] | null},
            
            \texttt{"targeted\_sexual\_orientations": list[str] | null},
            
            \texttt{"targeted\_genders": list[str] | null},
            
            \texttt{"targeted\_age\_groups": list[str] | null},
            
            \texttt{"targeted\_disabilities": list[str] | null},
            
            \texttt{"targeted\_professions": list[str] | null},
            
            \texttt{"targeted\_social\_classes": list[str] | null},
            
            \texttt{"others": list[str] | null}\}\\

        \textbf{Notes:}
        Do not include any explanation or additional information.\\
        \hline
    \end{tabular}
    \caption{Prompt for extracting hate targets from Telegram chats using Llama-3.1-Nemotron-70B-Instruct}

    \label{tab:identity_attack}
\end{table}